\documentclass{aa}  

\usepackage{graphicx}
\usepackage{txfonts}
\usepackage{color}
\usepackage{newtxtext,newtxmath}
\usepackage{graphicx}	
\usepackage{amsmath}	
\usepackage{amssymb}	
\usepackage{animate}
\usepackage{ulem}
\usepackage{hyperref}
\hypersetup{colorlinks=true,linkcolor=blue,citecolor=blue,filecolor=blue,urlcolor=blue}

\DeclareRobustCommand{\VAN}[3]{#2}
\let\VANthebibliography\thebibliography
\def\thebibliography{\DeclareRobustCommand{\VAN}[3]{##3}\VANthebibliography}

\begin{document} 

   \title{Photometric properties of classical bulge and pseudo-bulge galaxies at $0.5\le z<1.0$}
   \titlerunning{The classical and pseudo-bulge galaxies}
   \authorrunning{Hu et al.}


   \author{Jia Hu\inst{1,2}\thanks{E-mail: hujia@nao.cas.cn},
          Qifan Cui\inst{1,3}\thanks{qfcui@nao.cas.cn},
          Lan Wang\inst{1,2},
          Wenxiang Pei\inst{1,2,4}
          \and
          Junqiang Ge\inst{1,2}
        }
        
   \institute{National Astronomical Observatories, Chinese Academy of Sciences, Beijing 100101, China\
        \and
        School of Astronomy and Space Science, University of Chinese Academy of Sciences, Beijing 100049, China\
        \and
        Key Laboratory of Space Astronomy and Technology, National Astronomical Observatories, Chinese Academy of Sciences, Beijing 100101, China\
        \and
        Institute for Frontiers in Astronomy and Astrophysics, Beijing Normal University, Beijing 102206, China\
    }

   \date{Received 00 00, 2024; accepted 00 00, 2024}

 
  \abstract
   {We compare the photometric properties and specific star formation rate (sSFR) of classical and pseudo-bulge galaxies with $M_* \ge 10^{9.5} \rm M_{\odot}$ at $0.5\le z<1.0$, selected from all five CANDELS fields. 
   We also compare these properties of bulge galaxies at lower redshift selected from MaNGA survey \citep{2024MNRAS.529.4565H}.}
   {This paper aims to study the properties of galaxies with classical and pseudo-bulges at intermediate redshift, to compare the differences between different bulge types, and to understand the evolution of bulges with redshift.  
   }
   {Galaxies are classified into classical bulge and pseudo-bulge samples according to the S$\mathrm{\acute{e}}$rsic index n of the bulge component based on results of two-component decomposition of galaxies, as well as the position of bulges on the Kormendy diagram. 
   For the 105 classical bulge and 86 pseudo-bulge galaxies selected, we compare their size, luminosity, and sSFR of various components.}
   {At given stellar mass, most classical bulge galaxies have smaller effective radii, larger $B/T$, brighter and relatively larger bulges, and less active star formation than pseudo-bulge galaxies. 
   Besides, the two types of galaxies have larger differences in sSFR at large radii than at the central region at both low and mid-redshifts. 
   }
   {The differences between properties of the two types of bulge galaxies are in general smaller at mid-redshift than at low redshift, indicating that they are evolving to more distinct populations towards the local universe. Bulge type is correlated with the properties of their outer disks, and the correlation is already present at redshift as high as $0.5<z<1$.
   }


   \maketitle
%
\section{Introduction}
\label{sec:Intro}

In the central region of a spiral galaxy, there is often a bulge component that contains a higher concentration of stars and light compared to the disk component that extends to a larger radius \citep{1961hag..book.....S}.
Bulges are classified into two main categories based on their properties observed at low redshift.
Like elliptical galaxies, typical classical bulges are dominated by random motion and follow the fundamental plane \citep{2016ASSL..418...41F}.
In contrast, pseudo-bulges exhibit characteristics similar to disk galaxies, with rotational motion and notable substructures such as nuclear bars, spiral arms, and inner rings \citep{1997AJ....114.2366C,2002AJ....124...65E,2005A&A429.141K,2011ApJ...733L..47F}. 
In general, classical bulge galaxies are more massive, redder \citep{2007ApJ...664..640D,2010ApJ...716..942F,2019MNRAS.484.3865W,2020ApJ...899...89S,2022MNRAS.515.1175H}, and have larger central velocity dispersion \citep{2012ApJ...754...67F,2020ApJ...899...89S} than pseudo-bulge galaxies.

Theoretically, classical bulges are commonly attributed to violent processes like clumpy collapse \citep{1999ApJ...514...77N,2008ApJ...688...67E,2012MNRAS.422.1902I} and galaxy mergers \citep{2001A&A...367..428A, 2005A&A...430..115H}, while pseudo-bulges are thought to form slowly through secular evolution \citep{1996ApJ...457L..73C,2004ARA&A..42..603K,2005MNRAS.358.1477A}, involving mild gas inflow and stellar migration triggered by bars or spiral arms \citep{2010MNRAS.407L..41S,2012MNRAS.420..913B,2013ApJ...772...36G,2014MNRAS.439..623G,2015A&A...578A..58H,2020IAUS..353..155C,2022MNRAS.512.2537G}.
However, simulations, semi-empirical models \citep{2010ApJ...715..202H,2013ApJ...772...36G,2015A&A...579L...2Q,2018MNRAS.473.2521S,2016MNRAS.462L..41F} and observations \citep{2008ApJ...687...59G,2013A&A...552A..67E,2016ASSL..418...77L} indicate that bulge formation and evolution is more complicated than the general picture.

In the previous work of \citet{2024MNRAS.529.4565H}, with the help of the MaNGA survey \citep{2015ApJ...798....7B} and two-component decomposition catalogue \citep{2022MNRAS.509.4024D}, we compared the resolved properties of classical and pseudo-bulges and their host galaxies at $0.01<z<0.15$. 
We found that at given stellar mass, disks of pseudo-bulge galaxies are younger, have more active star formation, rotate more, and may contain more HI content compared with the disks of classical bulge galaxies.
More interestingly, the differences between the properties of disks for the two types of bulge galaxies are larger than the differences between bulges themselves, indicating that the formation of bulges may be closely related to the evolution of outer disks in galaxies.

While most statistical studies of bulge properties focus on the low redshift range, studying the bulge component of faraway galaxies is more difficult due to the limited resolution of observation.
The central region of higher-redshift galaxies can be resolved only by space telescopes, such as Hubble Space Telescope \citep[HST, ][]{polidan1991hubble}, James Webb Space Telescope \citep[JWST, ][]{2006SSRv..123..485G} and China Space Station Telescope \citep[CSST, ][]{2018cosp...42E3821Z}.
For example, \citet{2017ApJ...840...79S} explore the growth of different types of bulge galaxies since $z \sim 1$ using HST and find that classical bulges are brighter than pseudo ones at all redshift ranges. 
However, there are few studies about the other properties of bulges at mid-redshift.

In this work, to investigate more the properties of bulges and bulge galaxies at intermediate redshift, we study galaxies of different types of bulges at $0.5\le z<1.0$ from five CANDELS \citep{2011ApJS..197...35G,2011ApJS..197...36K,https://doi.org/10.17909/t94s3x} fields observed by HST. 
To select galaxies with bulge components, we perform two-component decomposition for the galaxies with $M_* \ge 10^{9.5} \rm M_{\odot}$. 
Following the previous work of \citet{2024MNRAS.529.4565H}, we divide selected samples into classical and pseudo-bulge galaxies using the criteria of bulge S$\mathrm{\acute{e}}$rsic index combined with the position of bulges on the Kormendy diagram \citep{2009MNRAS.393.1531G}.
We compare the photometric properties of bulges and disks including size and luminosity, as well as sSFR calculated by rest-frame UVI colour, and sSFR profiles for the two types of bulge galaxies.

This paper is organised as follows. 
Section \ref{sec: Sample selection} introduces how we select galaxies with different types of bulges from CANDELS. 
Section \ref{sec:Photometric} studies and compares the photometric properties of bulges and disks.
Section \ref{sec: UVI and sSFR} compares sSFR for the two types of bulges, their galaxies, and the sSFR profile of the galaxies. 
Conclusions and discussions are presented in Section \ref{sec:Summary}.
Throughout the paper we adopt flat $\Lambda$-CDM concordance model ($\rm{H_0=70.0 km}$ $\rm{s^{-1}}$ $\rm{Mpc^{-1}}$ and $\rm{\Omega_M=0.30}$).

\section{Two-component decomposition and sample selection}
\label{sec: Sample selection}

We perform two-component decomposition fitting for 2178 galaxies with good image quality selected from five CANDLES fields, and select 344 most reliable two-component galaxies from them. 
Then we classify galaxies with classical and pseudo-bulges by criteria of both the bulge S$\mathrm{\acute{e}}$rsic index and the position of bulges on the Kormendy diagram. 
In the end, 105 classical bulge galaxies and 86 pseudo-bulge galaxies are selected. 

\subsection{Selection of two-component galaxies}
\label{sec: Selection of two-component galaxies}

For all sources in the five catalogues of CANDELS fields, including COSMOS \citep{2017ApJS..228....7N}, EGS \citep{2017ApJS..229...32S}, GOODS-N \citep{2019ApJS..243...22B}, GOODS-S \citep{2013ApJS..207...24G} and UDS \citep{2013ApJS..206...10G}, we first select 2178 galaxies that have image quality good enough for further decomposition fitting, following the criteria presented in the previous works of \citet{2018ApJ...860...60L} and \citet{2024arXiv240717180C}. The details of selection criteria are listed below, and Tab. ~\ref{tab: samples number} presents the numbers of galaxies selected after applying each of the criteria.

$\circ$ CLASS\_STAR < 0.9 are sources identified as galaxies rather than stars, where CLASS\_STAR is an output index from Source Extractor \citep[\textsc{SExtractor},][]{1996A&AS..117..393B}. 

$\circ$ Only galaxies with F160W(H) band magnitude ($M_{H}$) brighter than 24.5 are selected, which have signal-to-noise ratios high enough to get reliable images.

$\circ$ GALFIT FLAG (H) = 0 selects galaxies that can be fitted reliably by GALFIT for F160W band images \citep{2012ApJS..203...24V}.

$\circ$ Galaxies with redshifts in the range of 0.5 $\le$ z < 1.0 are selected (see more detailed redshift measurements in \citet{2024arXiv240717180C}).

$\circ$ Small galaxies with $M_* < 10^{9.5} \rm M_\odot$ are removed.
The stellar mass is calculated by fitting SPS models to broadband photometry using the code \textsc{FAST} \citep{2009ApJ...700..221K}, based on models of \citet{2003MNRAS.344.1000B} with initial mass function of \citet{2003PASP..115..763C}, exponentially declining $\tau$-models of star formation history, solar metallicity, and a dust law of \citet{2000ApJ...533..682C}.

$\circ$ The criterion about semi-major axis of whole galaxies \citep[$\mathrm{R_{SMA}}\ge 0.18\arcsec$,][]{2012ApJS..203...24V} removes sources that are smaller than the full-width at half-maximum (FWHM) of point spread function (PSF) of HST images (average value in F160W band of 0.18$\arcsec$).

$\circ$ In the CANDELS catalogue, we select galaxies that are marked by SExtractor with PhotFlag = 0, indicating that they are not contaminated by foreground sources or neighboring objects, have no bad pixels, and are not located on the borders of the mosaic in the F160W band.

$\circ$ We have also visually inspected the images to ensure that the galaxies are uncontaminated in all bands of F160W, F606W, F814W, and F125W.

\begin{table}
	\centering
	\caption{Selection criteria and sample sizes for galaxies/sources in CANDELS.}
	\label{tab: samples number}
        \tabcolsep=1.0cm
        \renewcommand{\arraystretch}{1.3}
	\begin{tabular*}{1.0\columnwidth}{cc} 
		\hline
		Criterion &    Number\\
		\hline
		Full catalogue &    186435\\
            CLASS\_STAR < 0.9 &    179605\\
		$M_{H}$ $\le$ 24.5 &    110344\\
		GALFIT FLAG (H) = 0 &    39875\\
            0.5 $\le$ z < 1.0 &    12699\\
            $M_* \ge 10^{9.5}M_\odot$ &    4135\\
            $\mathrm{R_{SMA} \ge 0.18 \arcsec}$ &    3625\\
            PhotFlag = 0 &    2891\\
            Good multi-band data &    2178\\
		\hline
	\end{tabular*}
\end{table}

Combining the above criteria, we finally select 2178 galaxies to perform the two-component decomposition fitting using GALFIT.
Firstly, we conduct preprocessing for images of sources before the fitting.
We generate multi-band thumbnail images with a size of 401 × 401 pixels from the original CANDELS mosaics for individual targets. 
The sky background using a pixel-by-pixel background estimation method is subtracted. 
In addition, we take photometric parameters provided by \textsc{SExtractor} as initial structural parameter priors for \textsc{GALFIT} \citep{2002AJ....124..266P}.

Using \textsc{GALFIT}, we fit F160W images of our selected galaxies by two photometric function models: a single S$\mathrm{\acute{e}}$rsic profile \citep[Ser, ][]{1963BAAA....6...41S}, and a two-component model consisting of a S$\mathrm{\acute{e}}$rsic profile for the bulge and an exponential profile for the disk (SerExp).
During the fitting process, we set some limitations.
For both models, the range of index n on S$\mathrm{\acute{e}}$rsic profile is from 0.5 to 8 and the effective radius range is from 0.5 to 50 pixels. 
For SerExp models, we set the effective radii of disks to be larger than those of bulges ($R_b < R_d$), to ensure domination of disk components in the outer area.
As a result, we get the parameters of each galactic component for each photometric function model, such as effective radius, magnitude, axis ratio, and position angle.

\begin{figure*}
	\centering
	\includegraphics[width=2.0\columnwidth]{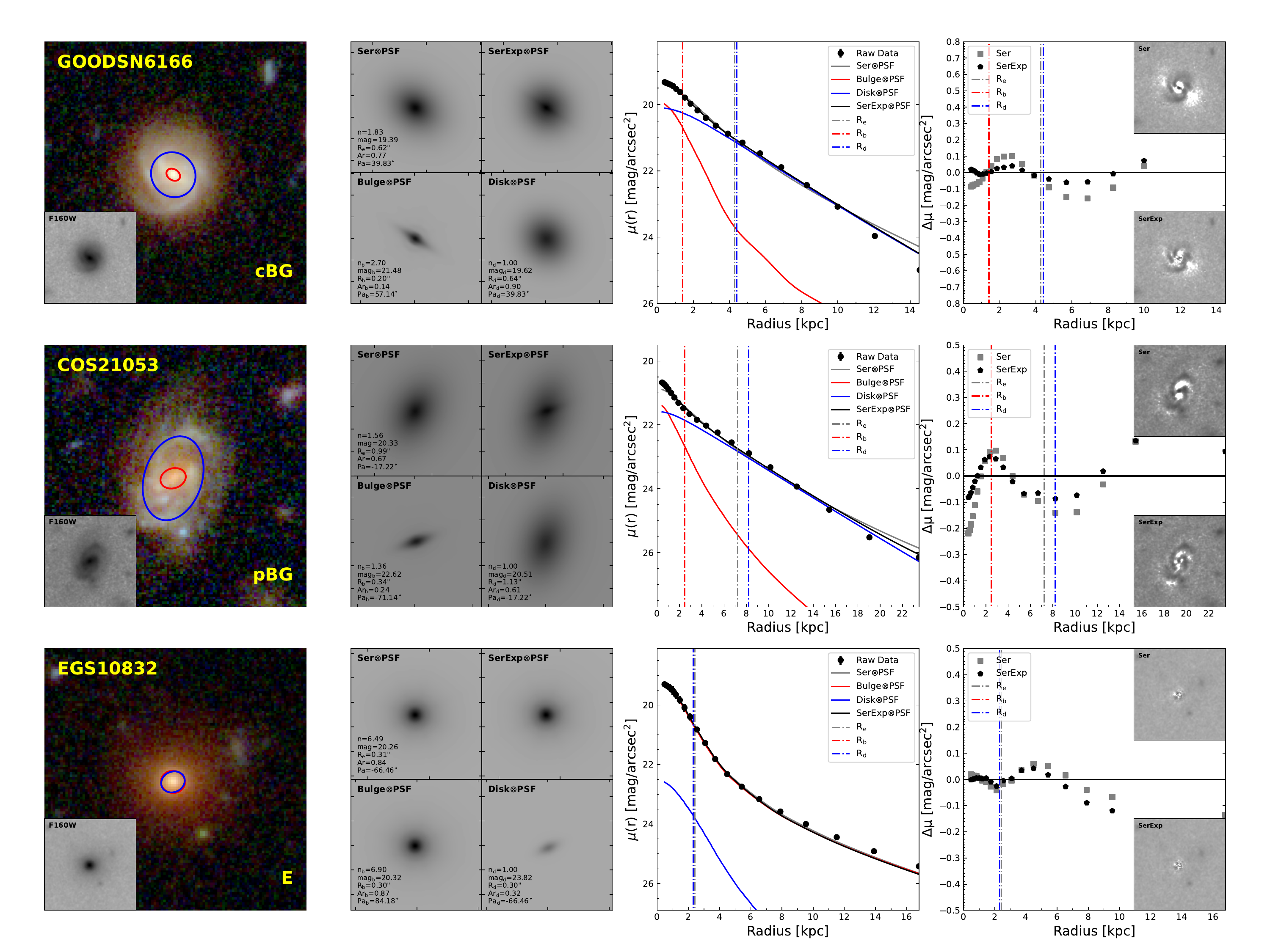}
	\caption{Two-component decomposition results of three example galaxies with different types. From top to bottom, a classical bulge galaxy (cBG), a pseudo-bulge galaxy (pBG), and an elliptical galaxy (E) are shown in each row. Their IDs and types are shown in the top left and bottom right corners in the most left panels. In each row, from left to right: (1) the first panel shows colour image and F160W image (inserted). The red and blue ellipses are the effective radii range of the bulge and disk components of SerExp fitting. (2) the second panel displays model images of Ser (top left sub-panel) and SerExp (top right sub-panel) models, where SerExp model image is the sum of model image of its bulge (bottom left sub-panel) and disk (bottom right sub-panel). The effect of PSF has been included in these model images. The outputs of fitting for each model are shown on the left bottom corners of each sub-panel, including S$\mathrm{\acute{e}}$rsic index (n), apparent magnitude (mag), effective radius (R), axial ratio (Ar), and position angle (Pa). (3) the third panel presents light profiles, where black dots are raw data, and the solid grey line displays Ser model fitting result. The solid black line is SerExp model fitting result, which is the sum of its bulge and disk represented by solid red and blue lines, where PSF is also included. Vertical grey, red, and blue dash-dot lines are the effective radii of Ser model $R_e$, bulge $R_b$, and disk $R_d$. (4) the fourth panel shows the 1D and 2D residuals. Grey square and black pentagon are the 1D residuals for Ser and SerExp models respectively. The 2D residuals for Ser and SerExp are displayed on top and bottom right corners.}
    \label{fig: image}
\end{figure*}

Based on the fitting results of GALFIT, we select the most robust two-component galaxy samples for which the SerExp model is preferred. In Fig.~\ref{fig: image}, we show examples of three galaxies that are identified as classical bulge + disk, pseudo-bulge + disk, and one-component elliptical by our method. The robust two-component galaxies are required to follow all the selection criteria below strictly: 

$\circ$ According to both F160W image and colour image (combined by F606W, F814W, and F160W images), 778 galaxies are selected in which a brighter and more concentrated central component can be visually identified, together with an extended bluer component in the outer region with some structure, like spiral arms or star formation clumps. 
For example, as shown in the leftmost three panels of Fig.~\ref{fig: image}, the top two galaxies exhibit brighter central components and bluer disks with spiral arms, while the bottom one shows a single smooth component.

$\circ$ For a two-component galaxy, its bulge light is required to dominate in the inner region and its disk to dominate in the outer region by analyzing the light profiles of different model fitting results. The one-dimensional residuals, representing the difference between real data and model data, must be smaller than $\pm 0.1$ at all radii. For example, in the third panel of the third column of Fig.~\ref{fig: image}, the light profile shows that its disk does not dominate in the outer region; thus, it is classified as a single-component galaxy. Based on these criteria, we select 657 galaxies for which the SerExp model is preferred.

$\circ$ The two-dimensional SerExp model images of the bulge/disk components reproduce well bulges/disks in the F160W images, showing similar effective radius, axis ratio, and position angle visually between them. 
This also means that the relative residuals from the two-dimensional SerExp modal data must be closer to 0 at all pixels.
For example, in the second and fourth columns of the first and second row in Fig.~\ref{fig: image}, the SerExp model results give good fitting to the images, with much smaller two-dimensional residuals.
Therefore 187 galaxies with inadequate results of SerExp model are discarded.

$\circ$ Robust bulge components are required to have effective radii greater than the half-width at half-maximum (HWHM) of PSF of HST images \citep{2009MNRAS.393.1531G}, which has an average value in F160W band of 0.09$\arcsec$. 
126 galaxies with small bulges are removed.

Finally, 344 bulge+disk galaxies with reliable fitting results are selected to be analysed in this work.
In addition, we also select elliptical galaxies to set the Kormendy relation criterion for identifying bulge types in the following subsection. 
Similar to the example shown in the third row of Fig.~\ref{fig: image}, 392 elliptical candidates best-fitted by only one smooth component on F160W are selected.
Combined with large S$\mathrm{\acute{e}}$rsic index ($3<n<7.95$), 253 typical elliptical galaxies are selected.

We choose the F160W band to be studied and presented in this paper, because the images have higher signal-to-noise ratio than the other bands. 
Besides, F160W band is relatively close to the r-band in SDSS at rest-frame, which is generally considered the good band where bulge features are most prominent observationally\citep{2016ASSL..418...41F}.
We have also performed the same fitting procedures on images in other bands and have checked that the obtained classification results of galaxy types remain similar.

\subsection{Selection of classical and pseudo-bulge galaxies}
\label{sec: Selection of classical and pseudo-bulge galaxies}

For classifying bulge types in low redshift galaxies, there are two common photometric criteria to divide bulge types \citep{2004ARA&A..42..603K,2016ASSL..418...41F,2019MNRAS.484.3865W,2020ApJ...899...89S}. The first is the S$\mathrm{\acute{e}}$rsic index of bulge component $n_b$, because most classical bulges are found to have $n_b>2$ and most pseudo-bulges have $n_b<2$ \citep[e.g.][]{2004ARA&A..42..603K,2016ASSL..418...41F}. The other criterion is the positions of bulges on the Kormendy diagram \citep{1977ApJ...218..333K}, where classical bulges are found to follow the Kormendy relation between effective radius $R_e$ and average surface brightness within the effective radius $\langle \mu_e \rangle$ of elliptical galaxies, while pseudo-bulges normally deviate \citep{2009MNRAS.393.1531G}.
For mid-redshift galaxies, only a few works have been done \citep{2017ApJ...840...79S,2018MNRAS.478...41S}.
We visually inspect the central parts of galaxies and find that bulges with structures such as inner ring, nuclear bar, and inner spiral mostly also have small $n_b$ and are located deviating from the Kormendy relation. 
Therefore we select our mid-redshift classical and pseudo-bulge samples also using criteria of $n_b=2$ combined with the Kormendy relation.

For the 344 two-component galaxies selected as described in the previous subsection, following the classification method of \citet{2024MNRAS.529.4565H}, we first use $n_b=2$ as the divider to classify bulge types, and get 181 classical and 155 pseudo-bulge galaxies for our sample.
In our previous works \citep{2019MNRAS.484.3865W, 2024MNRAS.529.4565H}, we select classical bulge galaxies with $3<n_b<7.95$ at low redshift, to minimise the effect of uncertainty in fitting bulge S$\mathrm{\acute{e}}$rsic index.
In this work, to have a larger sample, we use $2<n_b<7.95$ to select classical bulge samples.

\begin{figure}
	\includegraphics[width=\columnwidth]{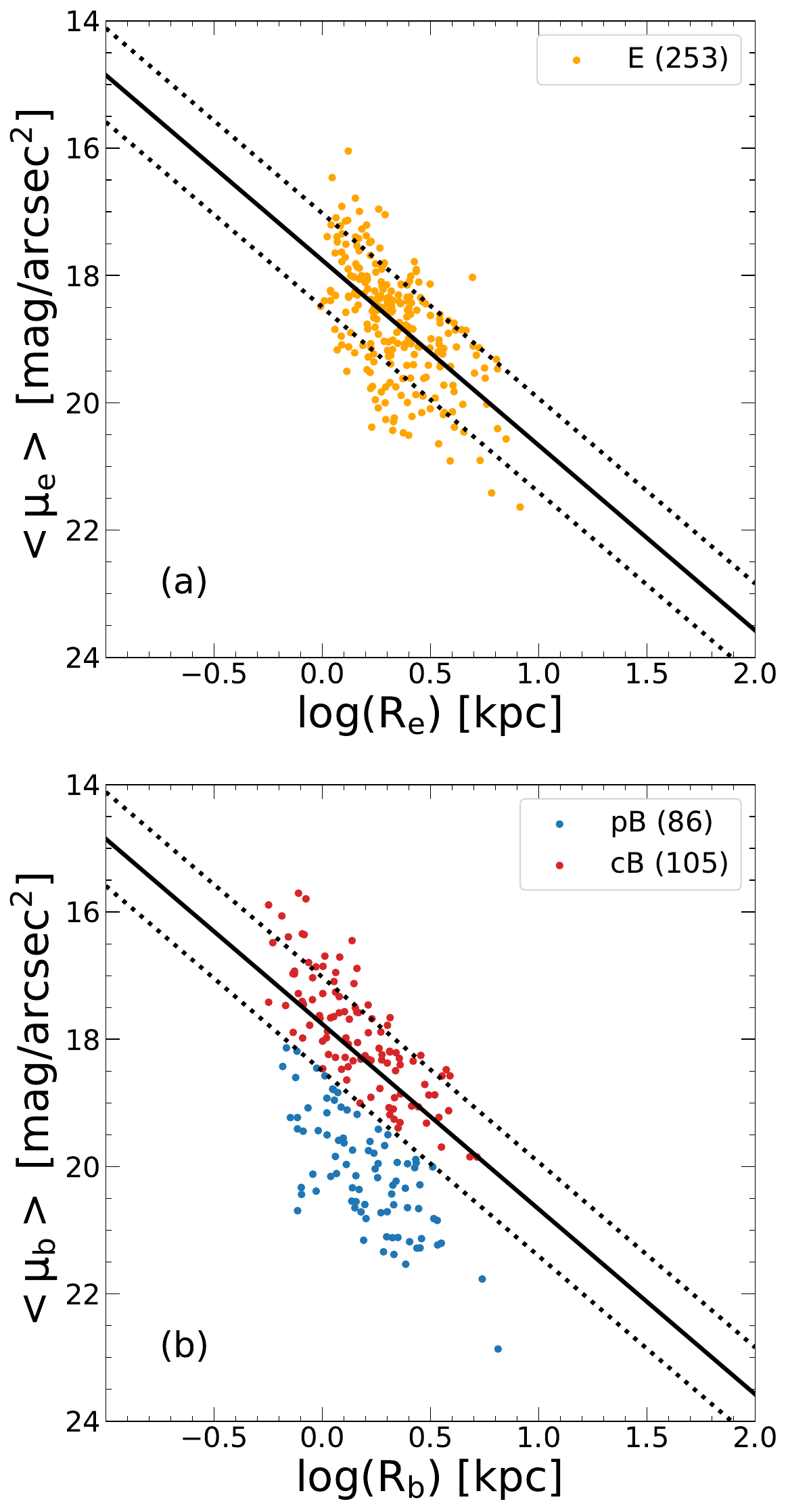}
    \caption{Panel (a): Kormendy Relation of our selected elliptical galaxies (E) in F160W band. The orange dots represent individual elliptical galaxies, and the black solid line is the best linear fitting of these dots. The two black dotted lines show the $\pm 1\sigma$ scatter. Panel (b): On the Kormendy diagram, the positions of classical bulges (cB: red dots) and pseudo-bulges (pB: blue dots) constrained by both $n_b$ and Kormendy relation are plotted.}
    \label{fig:kormendy relation}
\end{figure}

For the criterion of positions on the Kormendy diagram, we first fit the Kormendy relation of elliptical galaxies based on the 253 ellipticals selected as described in Section \ref{sec: Selection of two-component galaxies}. Panel (a) of Fig.~\ref{fig:kormendy relation} shows the result. The black solid line is the best linear fitting of these elliptical galaxies: 
\begin{equation}
    \langle \mu_e \rangle  = 2.91 \times \mathrm{log}(R_e/\mathrm{kpc})+17.76.
    \label{eq:eq3}
\end{equation}
The two black dotted lines show the $ 1\sigma$ scatter ($ \sigma$=0.74). 
We require the pseudo-bulges to lie below the lower dotted line in panel (a) of Fig.~\ref{fig:kormendy relation}, and the classical bulges to lie above the lower dotted line. 
Combined with the criterion of S$\mathrm{\acute{e}}$rsic index, we get 86 pseudo-bulges and 105 classical bulges, and their positions on the Kormendy diagram are indicated by blue and red dots in panel (b) of Fig.~\ref{fig:kormendy relation}.

\begin{figure}
	\centering
	\includegraphics[width=\columnwidth]{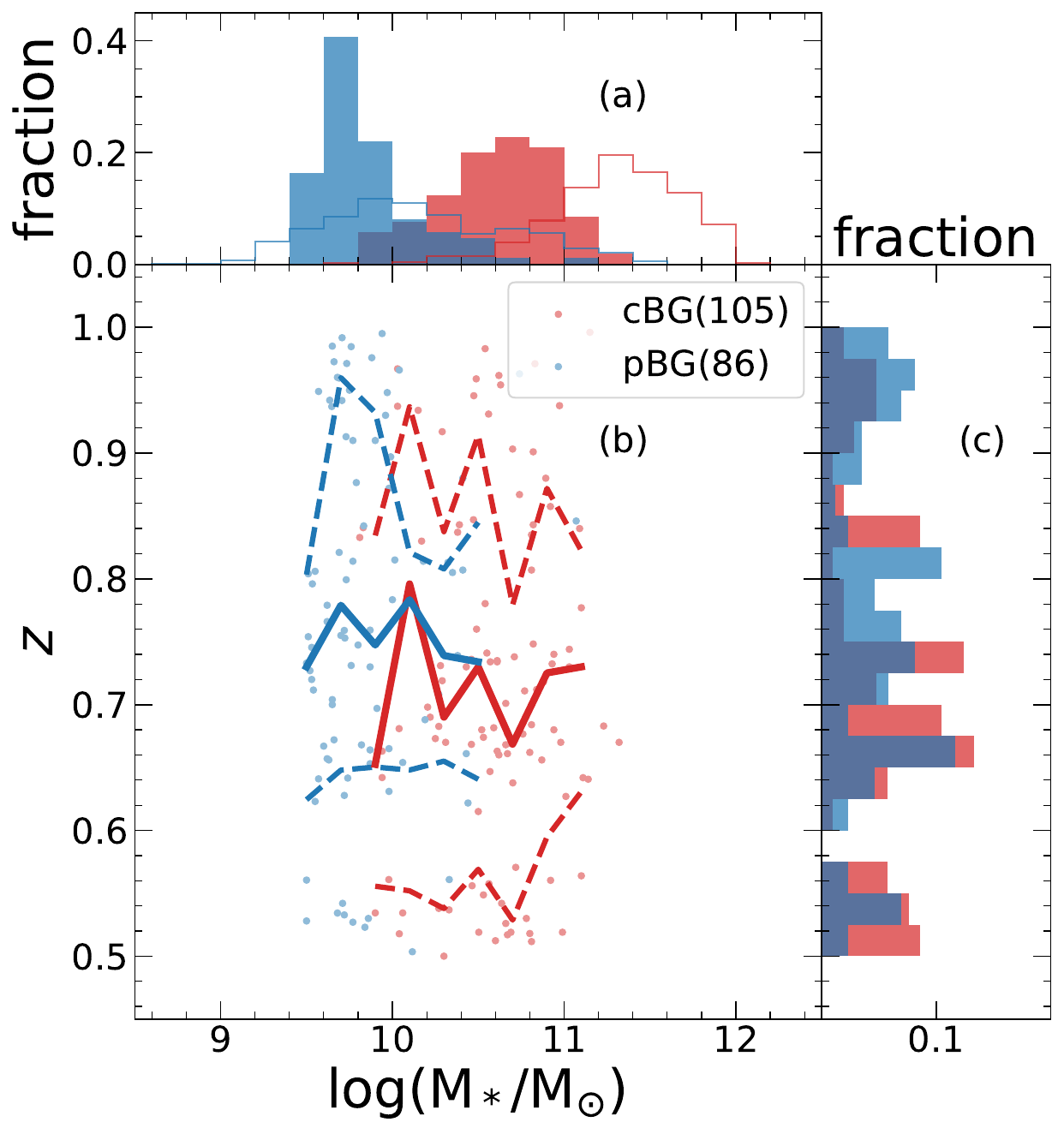}
	\caption{The distributions of galaxy total stellar mass and redshift for our selected classical bulge galaxies (cBG: red symbols) and pseudo-bulge galaxies (pBG: blue symbols). Panel (b) presents redshift as a function of galaxy total stellar mass. Each dot indicates an individual galaxy. Solid lines show median relations and dashed lines enclose 68\% scatter. Panels (a) and (c) present the distributions of the two properties respectively. In panel (a), unfilled histograms are over-plotted which show distributions of samples at low redshift from the previous work of \citet{2024MNRAS.529.4565H}.}
    \label{fig: fraction}
\end{figure}

In Fig.~\ref{fig: fraction}, we compare stellar mass and redshift of the selected galaxies with different types of bulges.  
In panel (a), the filled bars show the stellar mass distribution for classical and pseudo-bulge galaxies, and their peak values are $10^{10.7} \rm M_{\odot}$ and $10^{9.7} \rm M_{\odot}$ respectively.
For comparison, the unfilled bars are over-plotted, which show results of galaxies at low redshift from our previous work \citep[panel (a) of Figure 2;][]{2024MNRAS.529.4565H}. The peak stellar masses are $10^{11.3} \rm M_{\odot}$ and $10^{9.9} \rm M_{\odot}$ for classical and pseudo-bulge galaxies.
The comparisons indicate that both types of bulge galaxies at low redshift are more massive than galaxies at mid-redshift, and classical bulge galaxies are more massive than pseudo-bulge galaxies at all redshifts investigated.
Panels (b) and (c) of Fig.~\ref{fig: fraction} show that redshifts of two types of bulges both have wide distributions, independent of stellar mass.

\section{Photometric properties of bulges and bulge galaxies}
\label{sec:Photometric}

In this section, we compare the size and absolute magnitude of the selected classical and pseudo-bulges, and those of their host galaxies, as a function of galaxy stellar mass. 
The results are also compared with the properties of bulges and bulge galaxies studied in our previous work at low redshift \citep{2024MNRAS.529.4565H}, to get a clue of the evolution of bulges with cosmic time.

Note that bulge samples selected in this work from CANDELS and those selected at low redshift by \citet{2024MNRAS.529.4565H} are obtained by two-component fitting in bands with different wavelengths. 
F160W band is used for mid-redshift samples from the CANDELS survey, and the corresponding rest-frame wavelength is 8000 Å for galaxies at $z=1$. For the low redshift MaNGA sample, we use the SDSS $r$-band (6165 Å) for analyses.
Caution needs to be exercised when interpreting the comparison between the two.

\subsection{Size of bulges and bulge galaxies}
\label{sec: size}

\begin{figure*}
	\centering
	\includegraphics[width=1.34\columnwidth]{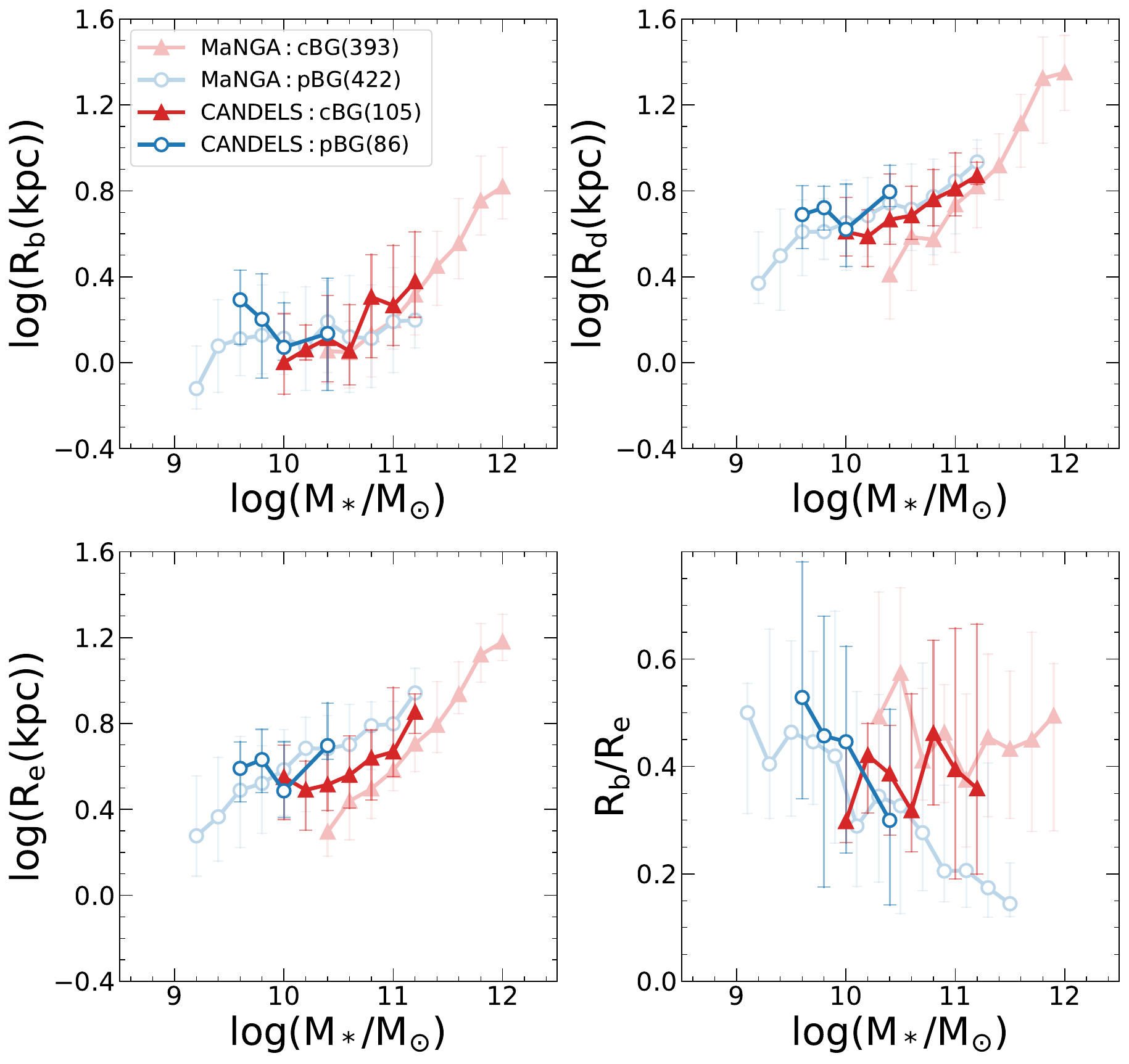}
	\caption{Sizes of bulges and bulge galaxies. Effective radii of bulges ($R_b$, top left panel), disks ($R_d$, top right panel), and whole galaxies ($R_e$, bottom left panel) as a function of galaxy stellar mass. Ratios of bulge effective radius to the effective radius of the whole galaxy ($R_b/R_e$, bottom right panel) as a function of galaxy stellar mass. The median values are indicated by red triangles for classical bulge samples and blue circles for pseudo-bulge samples. The error bars show the $1\sigma$ scatter. Dark colour represents results of the selected samples from CANDELS in this work, and light colour represents results of the samples selected at low redshift from MaNGA by \citet{2024MNRAS.529.4565H}.}
    \label{fig: size}
\end{figure*}

In Fig.~ \ref{fig: size}, we compare the sizes of bulges of different types and the sizes of their host galaxies and disk components.
The effective radius of each component is adopted to represent the size, which is from the two-component decomposition fitting as described in section \ref{sec: Selection of two-component galaxies}.

The top left panel of Fig.~ \ref{fig: size} compares the sizes of bulges. 
For CANDELS galaxies, sizes of pseudo-bulges decrease at $M_*<10^{10} \rm M_{\odot}$, and become almost flat or with a slight increase at larger masses. 
For classical bulges, their sizes remain similar at $M_*<10^{10.6} \rm M_{\odot}$, and then increase with stellar mass.
Both types of bulges have similar sizes in the overlapped region of stellar mass. 
Compared with the low redshift bulge samples selected by \citet{2024MNRAS.529.4565H}, as indicated by light symbols and lines, the sizes of both types of bulges are similar at different redshifts, only with exception for pseudo-bulges in galaxies less massive than $\sim 10^{9.8} \rm M_{\odot}$.

In the top right of Fig.~ \ref{fig: size}, for both classical and pseudo-bulge galaxies, the disk effective radius ($R_d$) increases with stellar mass for $M_*>10^{10} \rm M_{\odot}$. 
Disks of pseudo-bulge galaxies are a bit larger than disks of classical ones with similar stellar mass. 
This general trend is consistent with low redshift samples, with a larger difference between bulge types at low redshift than at mid-redshift.
Compared with the top left panel of Fig.~ \ref{fig: size}, larger differences exist in disk size than in bulge size between classical and pseudo-bulge galaxies, at both low and mid-redshifts. 
For the effective radius of the whole galaxies as shown in the bottom left of Fig.~ \ref{fig: size}, a similar trend is seen in the radius of disks. 
The difference between galaxies with different bulge types is even larger, mainly due to a larger bulge-to-total ratio of classical bulge galaxies, as will be seen later in Fig. \ref{fig: magnitude}.

We check the relative size of bulges to the whole galaxies in the bottom right panel of Fig.~ \ref{fig: size}. 
For pseudo-bulge galaxies, the relative bulge size decreases significantly with stellar mass. 
For classical ones, the dependence on stellar mass is much smaller. 
In the overlapped mass range of $10^{10.2-10.4} \rm M_{\odot}$, pseudo-bulge galaxies have relatively smaller bulges than classical ones. 
The difference between two bulge type galaxies is much larger for low redshift samples than for mid-redshift samples, mainly due to smaller sizes of classical bulge galaxies at low redshift.

\subsection{Absolute magnitude and bulge-to-total ratio}
\label{sec: mag}

\begin{figure*}
	\centering
	\includegraphics[width=2.0\columnwidth]{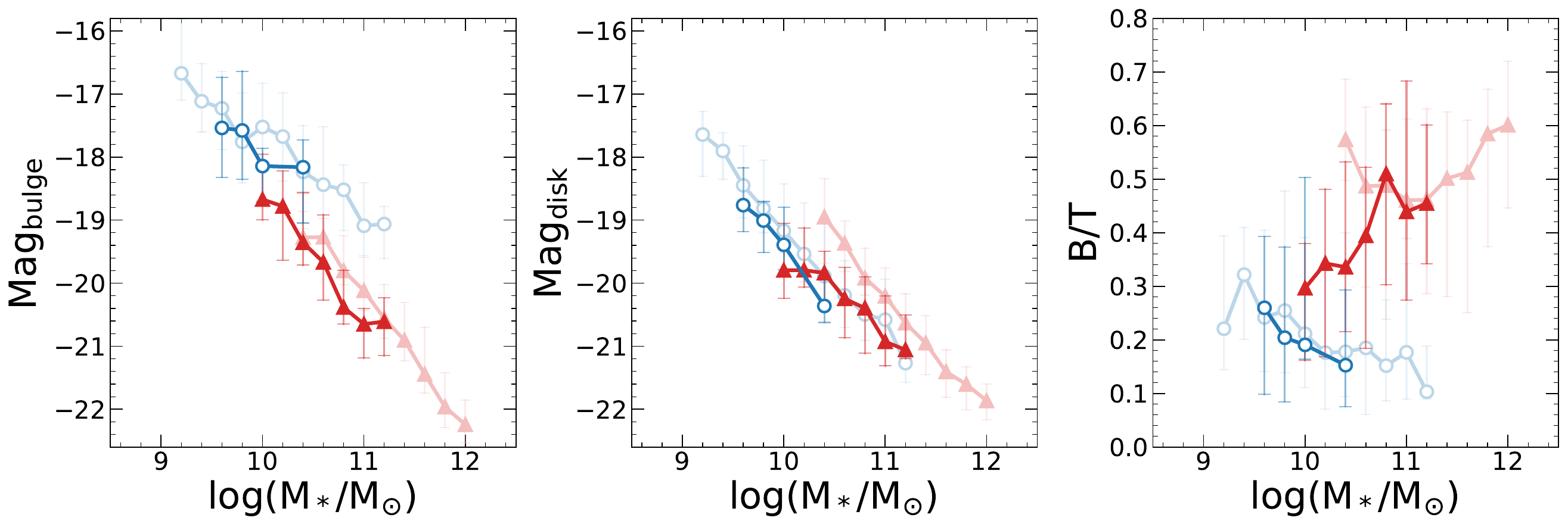}
	\caption{Absolute magnitude of bugle components (left panel), disk components (middle panel), and the bulge-to-total luminosity ratio($B/T$, right panel) as a function of stellar mass. The symbols and lines represent the same samples as in Fig.~\ref{fig: size}.}
    \label{fig: magnitude}
\end{figure*}

We compare in Fig.~ \ref{fig: magnitude} the absolute magnitude of our selected classical and pseudo-bulges, the absolute magnitude of disks of their host galaxies, and the luminosity bulge-to-total ratio ($B/T$) of their host galaxies.
The absolute magnitude of bulges and disks are from the results of two-component decomposition as described in section \ref{sec: Selection of two-component galaxies}. 
$B/T$ is the ratio between bulge luminosity and total luminosity of both bulge and disk components.

The left panel of Fig.~ \ref{fig: magnitude} shows that for galaxies both at mid- and low redshifts, bulges are brighter in more massive galaxies. 
In addition, classical bulges are brighter than pseudo-bulges in galaxies at given stellar mass at both mid- and low redshifts.
The result is consistent with \citet{2017ApJ...840...79S}, which shows that classical bulges are brighter than pseudo-bulges in B-band and I-band at $0.4 \le z<1.0$ and $0.02 \le z<0.05$.

In the middle panel of Fig.~ \ref{fig: magnitude}, with stellar mass increasing, the absolute magnitude of disks is brighter, at both mid- and low redshifts.
The absolute magnitude of the disks in the two types of galaxies is more similar at mid-redshift than at low redshift, with obviously brighter disks in pseudo-bulge galaxies than in classical bulge galaxies at low redshift. 
The results are also consistent with ones studied in \citet{2017ApJ...840...79S} in B- and I-band.

In the right panel of Fig.~ \ref{fig: magnitude}, we compare the $B/T$ for classical and pseudo-bulge galaxies as a function of galaxy stellar mass. 
At mid-redshift, pseudo-bulge galaxies have obviously smaller bulge fractions than classical ones at given stellar mass. 
Similar results have been derived by \citet{2018MNRAS.478...41S} when comparing $B/T$ of classical and pseudo-bulge galaxies in B-band at $0.4<z<1.0$.
At low redshift, the difference of $B/T$ between classical and pseudo-bulge galaxies is even larger. 
In particular, for classical bulge galaxies at stellar mass less than $10^{10.8} \rm M_{\odot}$, $B/T$ increases with stellar mass at mid-redshift while decreases with stellar mass at low redshift. 
The large $B/T$ of classical bulge galaxies leads to smaller effective radii of the galaxies, and makes the difference in $R_e$ more prominent than the difference in $R_d$ between pseudo-bulge and classical bulge galaxies, as seen in Fig. \ref{fig: size}.

\section{The sSFR of bulges and bulge galaxies}
\label{sec: UVI and sSFR}

In this section, we study and compare spatially resolved sSFR for the selected galaxies with classical and pseudo-bulges. 
For galaxies at mid-redshift from CANDELS, sSFR of each pixel in a given galaxy is estimated based on the position of galaxies on the rest-frame $UVI$ colour diagram, as shown in the top right panel of Fig. 10 in \citet{2024arXiv240717180C} and \citet{2017MNRAS.469.4063W}.
This new method is used to determine sSFR values, which have been tested to be consistent with results from broadband stellar population fitting from UV to Infrared \citep{2018ApJ...858..100F}.
Details regarding estimating rest-frame $UVI$ colour and sSFR can be found in Appendix A and B of \citet{2024arXiv240717180C}.

\begin{figure*}
	\centering
	\includegraphics[width=1.34\columnwidth]{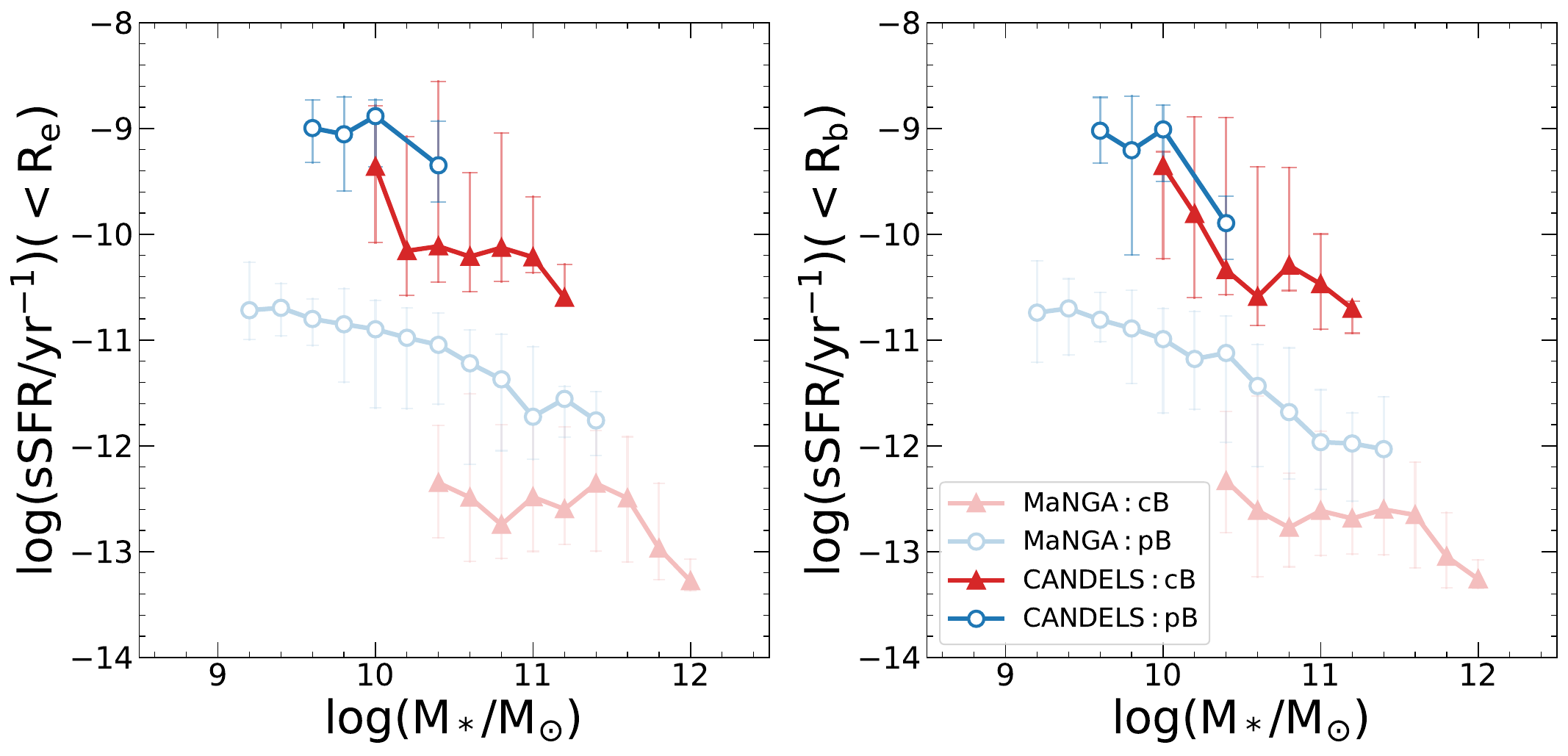}
	\caption{For galaxies with classical bulges and pseudo-bulges, sSFR within $R_e$ (left panel) and $R_b$ (right panel) as a function of galaxy stellar mass. The symbols and colours represent the same samples as in Fig.~\ref{fig: size}.}
    \label{fig: ssfr}
\end{figure*}

\begin{figure}
	\centering
	\includegraphics[width=1.0\columnwidth]{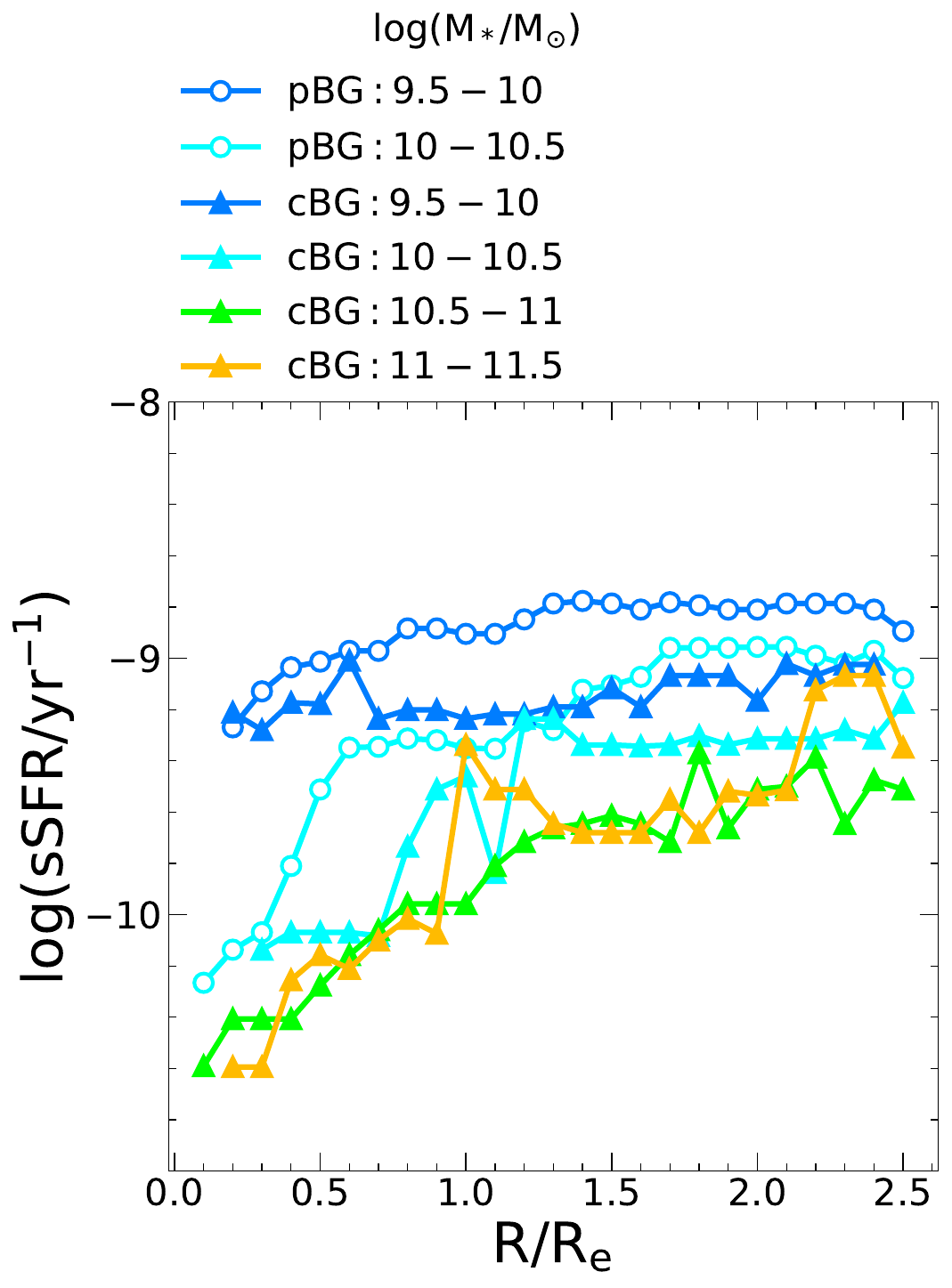}
	\caption{The sSFR radial profiles of selected classical (triangles) and pseudo-bulge galaxies (circles). Results are shown in galaxy stellar mass intervals of 0.5 in logarithmic space as indicated on the top of the panel using different colours.}
    \label{fig: profile}
\end{figure}

For comparison, we also show sSFR for bulge galaxies at low redshift in MaNGA selected by \citet{2024MNRAS.529.4565H}. 
SFR on each pixel of these galaxies is calculated according to $\rm H \alpha$ flux by empirical law of \citet[][equation 2]{1998ARA&A..36..189K}, where flux of $\rm H \alpha$ and stellar mass on each pixel of these MaNGA galaxies are from full spectrum fitting \citep{2004PASP..116..138C,2017MNRAS.466..798C,ge2018,2019MNRAS.485.1675G,2021MNRAS.507.2488G}.
As we will see in Fig.~\ref{fig: ssfr} and Fig.~\ref{fig: profile}, there exists systematic difference, up to two orders of magnitude, in the sSFR derived from the different methods applied to low and mid-redshift galaxies.
About one order of magnitude difference in sSFR at given stellar mass can be due to the intrinsic difference between galaxies at higher redshift of $z=1$ and galaxies at $z=0$, as studied by \citet{2018MNRAS.477.1822M} and \citet{2019MNRAS.488.3143B}.
\citet{2018MNRAS.477.1822M} compare sSFR of the same galaxies estimated from different methods, and find that the difference in sSFR values from $\rm H \alpha$ and rest-frame $UVI$ diagram is small.
Therefore, the other one-order of magnitude differences in sSFR between galaxies at low and mid-redshifts may be caused by different IMFs, flux calibrations, and other possible effects of spectral processing techniques.

In the left panel of Fig.~\ref{fig: ssfr}, we compare the average sSFR within $R_e$ of classical and pseudo-bulge galaxies indicated by red and blue symbols/lines. 
This value can be considered to represent the strength of star formation activities of the whole galaxies, since the average values of sSFR within an ellipse with a given radius vary with increasing radii $R$ and gradually become stable at $R \ge R_e$.
For galaxies at both redshifts, we find that galaxies with pseudo-bulges have more active star formation than galaxies with classical bulges at given stellar mass.
In addition, the difference in sSFR between classical and pseudo-bulge galaxies at mid-redshift is much smaller than that of galaxies at low redshift.

For the bulge components, we compare the sSFR within $R_b$ to reflect the bulge star formation activities, and the results are shown in the right panel of Fig.~\ref{fig: ssfr}.
sSFR of classical bulges is a bit smaller than pseudo-bulges at given stellar mass at both mid- and low redshifts.
However, the difference between bulge types at mid-redshift is obviously smaller than that at low redshift, and is also smaller than the difference between galaxies that host different types of bulges as shown in the left panel. 
This may indicate an evolution trend of different types of bulges to somehow deviate more in star formation activities from $z =1$ to $z=0$.

In Fig.~\ref{fig: profile}, we check further the profiles of the sSFR of galaxies with classical and pseudo-bulge at mid-redshift, to investigate in more detail the star formation activities in different components of galaxies.
The median sSFR for each galaxy sample is shown as a function of radius. 
The influence of PSF on the spectrum is considered, and the innermost points of galaxies are chosen to be greater than their HWHM.
Looking at the blue symbols in Fig.~\ref{fig: profile}, for galaxies with stellar mass in the range $10^{9.5-10}\rm M_{\odot}$, sSFR profiles are in general flat, with slightly increasing for both types of bulge galaxies.
For galaxies more massive than $10^{10}\rm M_{\odot}$, the sSFR profiles increase with increasing radius. 
Comparing the two types of bulge galaxies within the same stellar mass range, the sSFR of pseudo-bulge galaxies is similar and becomes larger than the classical ones from the center to the outer regions.

Comparing the sSFR profiles of galaxies at mid-redshift as shown in Fig. \ref{fig: profile} with the results of low redshift galaxies as shown in the middle panel of Fig. 10 from \citet{2024MNRAS.529.4565H}, the difference between the two bulge type galaxies is smaller at mid-redshift than at low redshift, consistent with the trend seen in Fig. \ref{fig: ssfr}. 
Besides, the slopes of sSFR for galaxies at mid-reshift are basically monotonic, while at low redshift pseudo-bulge galaxies mostly have peak sSFR at radius of around $0.8-1.4R_e$. 
The evolution of sSFR slopes of pseudo-bulge galaxies may give hint on the co-evolution of the bulge and the disk components.

\section{Summary}
\label{sec:Summary}

In this paper, we compare photometric properties and sSFR of classical and pseudo-bulges and their host galaxies at $0.5\le z<1.0$ selected from the CANDELS survey. We also compare the results with those derived at low redshift, based on the samples selected from MaNGA survey in an earlier work of \citet{2024MNRAS.529.4565H}.
The galaxies with classical bulge and pseudo-bulge are selected based on the S$\mathrm{\acute{e}}$rsic index n of bulge component by two-component decomposition fitting, combined with the criterion of the position of bulges on the Kormendy diagram.

Both at mid- and low redshifts, classical bulge galaxies are more massive than pseudo-bulge galaxies. 
At given stellar mass, most classical bulge galaxies have smaller effective radii of all components, larger $B/T$, brighter and relatively larger bulges, and lower sSFR at all radii than pseudo-bulge galaxies. 
However, the differences in most properties between two types of bulge galaxies are in general smaller at mid-redshift than at low redshift, indicating that these two types of bulges and their galaxies are evolving to more distinct populations towards the local universe.

Besides, the difference in central sSFR between classical and pseudo-bulge galaxies is smaller than the difference at larger radii for mid-redshift galaxies, which is consistent with what was previously discovered in \citet{2024MNRAS.529.4565H}. 
Therefore, the co-evolution of bulges with their host galaxies and outer disks is already present at redshift as high as $0.5 \le z<1$.

While it is difficult to simulate the formation and evolution of galactic bulges precisely due to the numerical effect suffered \citep{2024MNRAS.532.2558Z},
in future works, we will study further the observational properties of bulge galaxies at higher redshift, based on analysis of two-component decomposition fitting for galaxies at $1.0 \le z<3.0$ selected from JWST within five CANDELS fields.

\begin{acknowledgements}
This work is supported by the National Natural Science Foundation of China (grant No. 11988101), the National SKA Program of China (Nos. 2022SKA0110200, 2022SKA0110201), the National Key Research and Development Program of China (No. 2023YFB3002500), the Strategic Priority Research Program of Chinese Academy of Sciences, (grant No. XDB0500203), and K.C. Wong Education Foundation. JG acknowledges support from the Beijing Municipal Natural Science Foundation (No. 1242032), the Youth Innovation Promotion Association of the Chinese Academy of Sciences (No. 2022056), and the science research grants from the China Manned Space Project.
\end{acknowledgements}

\bibliography{references}{}
\bibliographystyle{aa}

\end{document}